\documentclass[conference]{IEEEtran}

\usepackage[english]{babel}
\usepackage[utf8x]{inputenc}
\usepackage[T1]{fontenc} 

\usepackage[linesnumbered]{algorithm2e}
\SetKwInOut{Input}{Input}
\SetKwInOut{Output}{Output}

\usepackage[a4paper,top=3cm,bottom=2cm,left=3cm,right=3cm,marginparwidth=1.75cm]{geometry}

\usepackage{amsmath}
\usepackage{graphicx}
\usepackage[colorinlistoftodos]{todonotes}
\usepackage[colorlinks=true, allcolors=blue]{hyperref}

\title{A Self-adaptive Agent-based System for Cloud Platforms}
\author{\IEEEauthorblockN{Merzoug Soltane}
\IEEEauthorblockA{University of El-Oued LINFI\\
El-Oued, Algeria\\merzoug-soltane@univ-eloued.dz}
\and
\IEEEauthorblockN{Yudith Cardinale}
\IEEEauthorblockA{Dpto. de Computación y T.I\\ Universidad Simón Bolívar, Venezuela\\
ycardinale@usb.ve}
\and
\IEEEauthorblockN{Rafael Angarita}
\IEEEauthorblockA{LISITE Laboratory, ISEP Paris\\
Paris, France\\
rafael.angarita@isep.fr}
\and
\IEEEauthorblockN{Philippe Rosse}
\IEEEauthorblockA{UPPA, LIUPPA\\
Anglet, France\\
philippe.roose@univ-pau.fr
}
\and
\IEEEauthorblockN{Marta Rukoz}
\IEEEauthorblockA{
Université Paris-Dauphine\\ 
CNRS, Paris, France\\
mrukoz@dauphine.fr}
\and
\IEEEauthorblockN{Derdour Makhlouf}
\IEEEauthorblockA{
Tebessa University 
\\ Tebessa, Algeria\\
m.derdour@yahoo.fr
}
\and
\IEEEauthorblockN{Kazar Okba}
\IEEEauthorblockA{
 Biskra University
\\Biskra, Algeria\\
kazarokba@gmail.com
}

}

\begin{document}
\maketitle

\begin{abstract}
Cloud computing is a model for enabling 
on-demand network access to a shared pool of computing resources,
that can be dynamically allocated and released with minimal effort. 
However, this task can be complex in highly dynamic environments with various resources to allocate for an increasing number of different  users requirements.
In this work, we propose a Cloud architecture based on a multi-agent system exhibiting a self-adaptive behavior to address the dynamic resource allocation. This self-adaptive 
system follows a {\it MAPE-K} approach to reason and act, according to  QoS,
Cloud service information, and propagated run-time information, to detect QoS degradation and make better resource allocation decisions.
We validate our proposed 
Cloud architecture  by 
simulation.
Results show that it can properly allocate resources to reduce energy consumption, while satisfying the users demanded QoS.
\end{abstract}

\vspace{-0.1cm}
\section{Introduction}
\vspace{-0.1cm}


Cloud computing is the most flexible 
computing model to provide on-demand platforms for accessing custom resources and applications in the form of several types of services. 
Most common services in the Cloud are Platform as a Service (PaaS), Infrastructure as a Service (IaaS), Software as a Service (SaaS), Data as a Service (DaaS). 
The constantly increasing number of such services on the Cloud arises new challenges for ensuring their correct use,
while guarantying 
self-adaptivity property, especially to manage user requirements (e.g., ensure QoS)~\cite{Gutierrez-Garcia2013} and optimize the use of resources (e.g., by minimizing energy consumption~\cite{srikantaiah2008energy}). 
Taking into account these aspects, resource allocation becomes a complex task in highly dynamic environments, because of the increasing number of managed resources, as well as the increasing number of users 
with different requirements.

Existing resource allocation strategies lead to two unwanted situations: (i) the waste of resources when the demand is more than the capacity and demands cannot be attended, and (ii) the waste of resources when the demand is less than the capacity. In the first case, there are users not satisfied; in both cases, there are idle resources that are misused. A self-adaptive approach can direct the system to an ideal situation in which capacity is dynamically adapted to the demand of resources. 


The main aim of this work is to address the resource allocation in Cloud platforms, towards a better utilization of resources 
and satisfaction of users requirements, by offering a self-adaptive approach. We propose a Cloud computing architecture that benefits from the combination of the Cloud and multi-agent technologies. 
The self-adaptive multi-agent system integrated into the proposed Cloud architecture  is  inspired on our previous work presented in~\cite{Angarita2016WWW,Angarita2016}. This self-adaptive system is a  \textit{MAPE-$K$-based (monitoring, analyzing, planning, executing, and knowledge)} approach~\cite{Iglesia2015} to reason and act, according to  QoS (e.g., time, price),
Cloud service information, and propagated run-time information, to detect QoS degradation, failures, or unavailability and make better resource allocation decisions for all users requests.
The agents application knowledge for decision making 
comprises off-line precomputed global and local information, user QoS preferences, and propagated actual resource state information. We validate our proposed multi-agent based Cloud architecture  by prototyping and simulation.
Results show that it can properly allocate resources to reduce energy consumption, while satisfying the QoS demanded by users.



\vspace{-0.1cm}
\section{Related Work}
\label{sec:related}
\vspace{-0.1cm}

Self-adaptive and autonomous management of computing resources have been topics of research interest and development for many years. In this section, we present recent literature concerning self-adaptive systems and agent-based approaches for cloud computing.

\vspace{-0.2cm}
\subsection{Self-adaptive system in Cloud computing}

In~\cite{Kertesz:2014:ISA:2565361.2565422}, a self-adaptive approach for Service-Level Agreement (SLA) based service virtualization in 
Cloud environments is  proposed. The proposed architecture allows the interoperability of service executions in 
heterogeneous, distributed, and virtualized environments. 
Inter Cloud is a self-management and SLA handling approach proposed in~\cite{buyya}. 
It is a Cloud federation oriented provisioning environment, that offers just in-time, opportunistic, and scalable application services. The idea of Inter Cloud is to envision utility-oriented federated IaaS systems that are able to predict application service behavior for intelligent down- and up-scaling infrastructures. Though it addresses self-management and SLA handling, the unified utilization of other services
like PaaS and SaaS are not studied. These SLA-based works 
demonstrate that the combination of negotiation, brokering, and deployment 
using SLA-aware extensions and autonomic computing principles are suitable 
for achieving reliable and efficient 
service operation in distributed 
environments. However, its application in resource allocation has not been proved.  


An approach for self-adaptive and self-configurable CPU resource provisioning for virtualized servers using Kalman filter\footnote{Kalman filtering is an algorithm that uses a series of measurements observed over time and produces estimates of unknown variables that tend to be more accurate than those based on a single measurement alone.} is presented in~\cite{Kalyvianaki}. That work deals with the integration of a Kalman filter into feedback controllers to dynamically allocate CPU resources to virtual machines hosting server applications, creating a new resource management scheme. The novelty of this approach is the use of the Kalman filter to track the CPU usage and update the allocations accordingly. 
Another approach based on monitored data is described in~\cite{Wang:2018}, which consists on a self-adaptive Cloud monitoring approach with on-line anomaly detection. The main contribution of this work is the design of a self-adaptive monitoring framework, that can efficiently collect monitoring data from various distributed systems deployed in a Cloud computing environment. The authors focused on a correlation-based method to select key metrics representing other metrics.  The problem with these works relies on the limitation of their applications. They are not generic and applicable for  all kind of resource allocation.

Other works focus on QoS compliant~\cite{Evans,SPEZZANO2016512}. In~\cite{Evans}, it is proposed a distributed self-adaptive architecture based on the Edge Computing concept with container-based technologies, such as Docker and Kubernetes, to ensure QoS for time-critical applications. For each container, features of resources required for the host can be allocated upon monitoring data and operational strategies defined by end-users, application developer, and administrator. In~\cite{SPEZZANO2016512}, it is proposed a QoS-aware Cloud Service Selection approach using service clustering and self-adaptation to 
efficiently support service selection in which runtime changes in the services QoS are taken into account to adapt to changes. As our work, these studies focus on QoS satisfaction; however, they do not consider local and remote information for more accurate decisions.

Based on the {\it MAPE} model, concepts of autonomic computing for adaptive management of Grid Computing are  leveraged in~\cite{4539332}. 
However, this approach does not consider the deployment and virtualization mechanism in Cloud datacenters.


\vspace{-0.1cm}
\subsection{Agent-based Clouds}
\vspace{-0.1cm}

Agents are networked software entities that can do specific tasks on behalf of a user and have a degree of intelligence that allows them to perform parts of their tasks autonomously and to interact with their environment in a successfully way. Agents are characterized by important features such as autonomy, sociality, rationality, responsiveness, proactiveness, and mobility~\cite{Vinyals2011}. Some studies have been proposed to design Cloud platforms based
on multi-agent systems~\cite{al2015,Fortino2014,singh2015}.
In~\cite{Fortino2014}, it is proposed the integration of agent-based system and Cloud computing for smart objects. This approach is suitable to effectively model Cooperative Smart Objects (CSO). In particular, a CSO is a smart object able to sense, store, and interpret information. 
 In~\cite {singh2015}, it is presented a mechanism to provide dynamic load balancing for Clouds based on autonomous agents. The proposed mechanism is based on Ant mobile agents and whenever the load of a Virtual Machine (VM) reaches a threshold value, it initiates a search for a candidate VM from other datacenters, reducing in this way the allocation time. This work only considers the workload of VMs and does not consider the physical machines.
In~\cite{al2015}, a multi-agent system is proposed
to manage the Cloud resources, while taking into account the customers QoS requirements. In this work a VM migration occurs when its hosting physical machine is facing an overloading or under loading problem. This approach is the most similar to ours, however, our  approach is  autonomic self-adaptive,  allowing controlling the user requirements depending on the system and agents states.
 



To overcome the limitation of existing works in the field of self-adaptive model for Cloud computing, we propose a novel approach for the management of resources in Cloud platforms. Our approach is based on a multi-agent system implementing the {\it MAPE-K} model to enable self-adaptation. 
Particularly, we enhance the Cloud computing architecture with agent technologies to tackle the energy of consumption problem.


\vspace{-0.1cm}
\section{Agents for the Cloud}
\label{sec:agents}
\vspace{-0.1cm}

One of the foundations of autonomic computing is the {\it MAPE-K} 
model for building autonomous self-adaptive agents~\cite{Iglesia2015}. 
{\it MAPE-K}-based agents can sense diverse properties of the environment and make \textit{rational} decisions based on them. Such an agent can \textit{monitor} changes on the environment where it exists, \textit{analyze} the situation based on the sensed information, \textit{plan} what it is going to do next based on this situation and its own capabilities and options, and \textit{execute} the plan. All of these steps are supported by the 
\textit{knowledge} the agent has about the environment and itself.
This model is usually implemented as a loop to permanently sense the environment and adapt to changes.



\begin{figure}
\centering
   \includegraphics[scale=0.3]{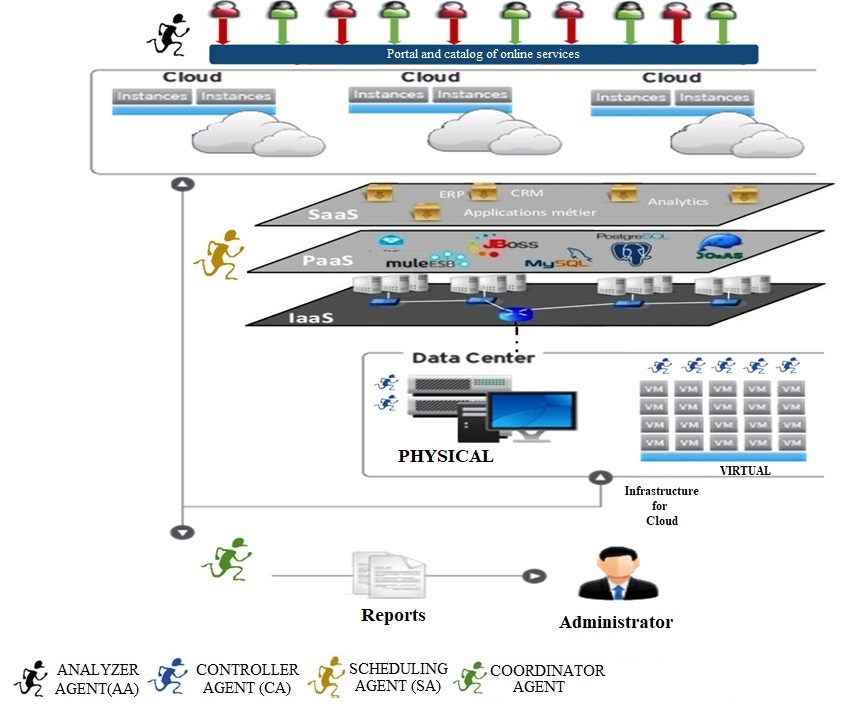}
   \vspace{-0.3cm}
   \caption{Agent-based Cloud Architecture.}
   \label{fig:CloudArchitecture}
   \vspace{-0.3cm}
\end{figure}

\begin{figure}
\centering
   \includegraphics[scale=0.3]{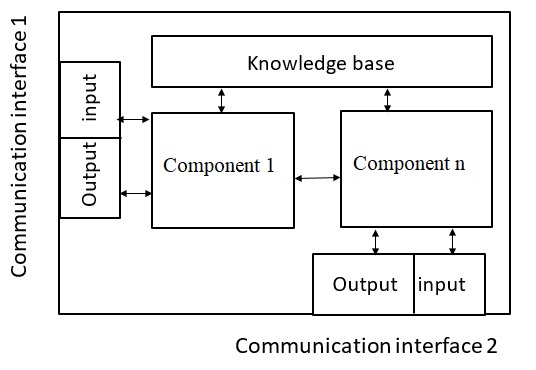}
   \vspace{-0.3cm}
   \caption{Agent Architecture Blueprint.}
   \label{fig:generalArchitecture}
   \vspace{-0.3cm}
\end{figure}

Figure~\ref{fig:CloudArchitecture} shows our proposed
general multi-agent Cloud Architecture,
based on three layers: user interface, Cloud instances, and  datacenter infrastructure. The user layer provides an interface to access the Cloud services and through which users specify their resource allocation needs. Analyzer Agents  identify  the resources and services demanded by users and build specific queries. Cloud instances layer represents an inter-layer between upper layer (i.e., users) and lower layer (i.e., datacenters). Scheduler Agents work at this layer. Datacenter layer provides resources as a service. It consists 
of a physical layer (e.g., servers, host, physical manage)  and a virtual layer (i.e., instances of VMs). 
There exists a Controller Agent per each physical machine in datacenters. Additionally, the Coordinator Agent supervises the whole process. Figure~\ref{fig:generalArchitecture} shows our proposed  agent architecture blueprint. Internal agent components rely on a modular architecture, favoring the 
understanding 
and 
reuse of the main components of each agent. 
Each component focuses on a specific function required for accomplishing the 
purpose of the agent. Each agent can  communicate with other agents through its communication interfaces. 
In this section we detail the analyzer agent, the scheduling agent, and the controller agent. 
The coordinator agent supervises the whole process.

\vspace{-0.1cm}
\subsection{The Analyzer Agent}

The analyzer agent is the first agent launched by the system. We illustrate its detailed architecture in Figure~\ref{fig:AAArchitecture}. Its goal is to study a user requirement (done by the Request Analyzer Component) and to build  a request specifying the resources needed to satisfy it (through the Resources Requester component). This agent 
relies on a knowledge base containing individual knowledge and analytical knowledge, as follows:

\begin{figure}
\centering
   \includegraphics[width=\columnwidth]{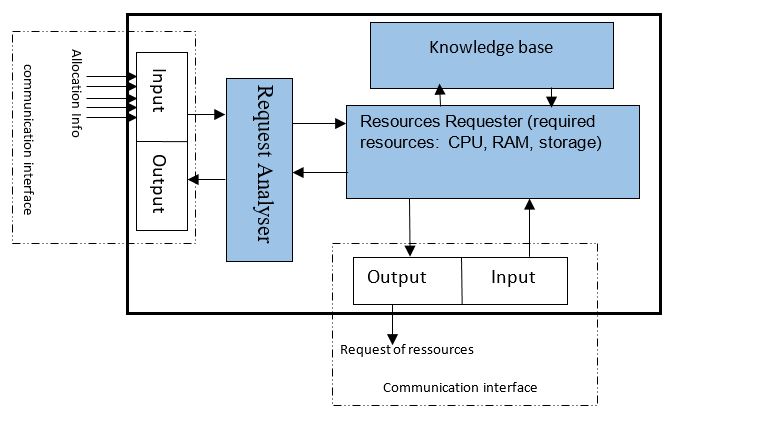}
   \vspace{-0.5cm}
   \caption{Analyzer Agent Architecture.}
   \label{fig:AAArchitecture}
   \vspace{-0.3cm}
\end{figure}

\noindent
{\bf - Individual knowledge:} it reflects the agent self-knowledge, including the following:
\begin{itemize}
\item \textit{name}: Agent analyzer (AA);
\item \textit{address}: the location where it is deployed; e.g., web Cloud service interface;
\item \textit{individual objectives}: its objective is analyze the resources requested by a user;
\item \textit{state}: it can have the following execution states: receiving user request, 
sending resource allocation request, and getting resource allocation response to/from the Scheduling Agent.
\end{itemize}

\noindent 
{\bf - Analytical knowledge:} it concerns rules stored in its knowledge base dictating its behavior and actions.

Its detailed behavior is illustrated in the algorithm presented in Alg.~\ref{alg:analyzerAgent}. It receives the user requirement, which is analyzed by the Request Analyser component (lines 2 to 4 in Alg.~\ref{alg:analyzerAgent}), by identifying the required resources (line 5 in Alg.~\ref{alg:analyzerAgent}). This information is received by the Resources Requester module (line 9 in Alg.~\ref{alg:analyzerAgent}), which in turn build a request with  the required resources to send it to the Scheduling Agent (lines 10 and 11 in Alg.~\ref{alg:analyzerAgent}).

\begin{algorithm}
  \scriptsize
 \Input{(i), service description; (ii), SAL; (iii), id. user}
 \Output{(i), needed resources}
 \Switch{communication interface}{
  \Case{input interface}{
  	\If{service description}{
    	analyzerProcess(service description)\;
        \textbf{return} resources = \{ CPU, RAM, disk,... \};
    }
  }
   \Case{output interface}{
   		\If{analyzerProcess returns resources}{
        	request = resources + id.user\;
            \textbf{send} request to the scheduling agent\;
        }
   }

 }
 \caption{Analyzer Agent (AA)}
 \label{alg:analyzerAgent}
 \vspace{-0.1cm}
\end{algorithm}

\vspace{-0.1cm}
\subsection{The Scheduling Agent}

\begin{figure}
\centering
   \includegraphics[width=\columnwidth]{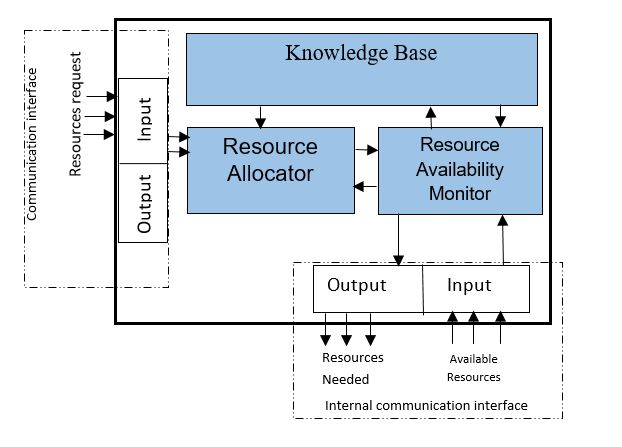}
   \vspace{-0.3cm}
   \caption{Scheduling Agent Architecture.}
   \label{fig:SAArchitecture}
   \vspace{-0.3cm}
\end{figure}

The Scheduling Agent is the central agent in the system. Its goal is to allocate resources needed by users. 
We illustrate its architecture in Figure~\ref{fig:SAArchitecture}. The Resource Allocator module receives requests from the Analyzer Agent and makes allocation decisions respecting the requested  resources and QoS, and according the available resources. The Resource Availability Monitor is in charge of updating the resources state.

The knowledge base of the scheduling agent is 
composed by two parts: 

\noindent 
{\bf - Individual knowledge:} reflects the agent self-knowledge, including:
\begin{itemize}
\item \textit{name}: Agent Scheduling (SA);
\item \textit{address}: the location where it is deployed; e.g., the resource management system in the data center;
\item \textit{individual objectives}: its objective is the resource allocation;
\item \textit{state}: this agent can have the following execution states: receiving request from the Analyzer Agent, sending a request of available resources, and waiting for a response to that request.
\end{itemize}

\noindent 
{\bf - Allocation knowledge:} 
\begin{itemize}
\item \textit{Available resources}: the available resources on all physical machines;
\item \textit{Self-adaptive options}: for example, retry, resource replacement or reallocation.
\item \textit{Inference engine and rules}: the inference engine applies rules to the knowledge base to deduce new information. Rules can specify the actions to take given certain conditions. 
\end{itemize}

\begin{figure}
\centering
   \includegraphics[width=\columnwidth]{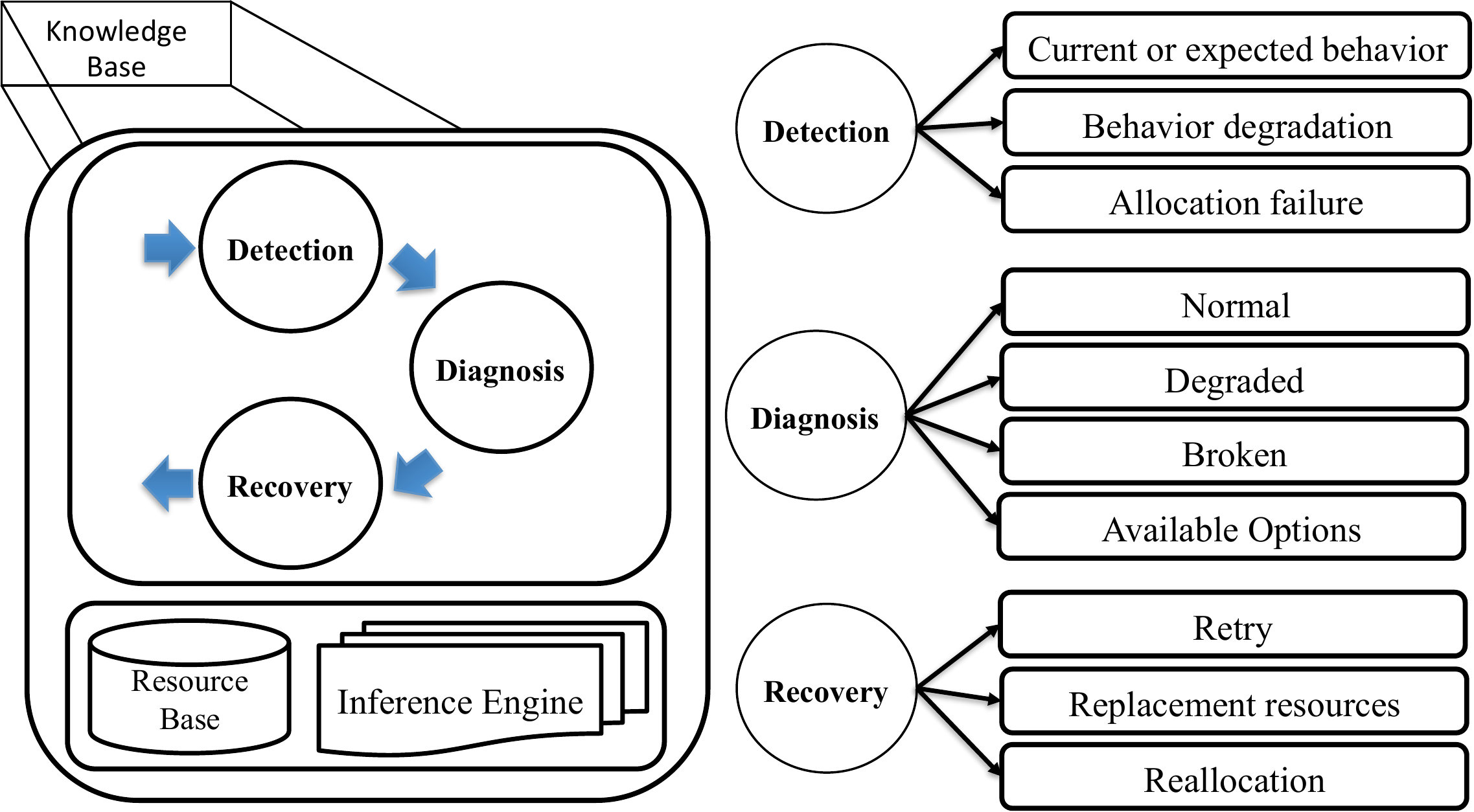}
   \vspace{-0.3cm}
   \caption{Scheduling Agent Behavior (General Description).}
   \label{fig:SAkb}
   \vspace{-0.4cm}
\end{figure}

Its detailed behavior is illustrated in Figure~\ref{fig:SAkb} and in the Alg.~\ref{alg:schedulingAgent}. If it receives a resource request, it checks the availability of those resources and sends an allocation request to the Controller Agent (lines 3 to 5 in Alg.~\ref{alg:schedulingAgent}). When it receives from the Controller Agent the availability of resources, it updates the information in its knowledge base and does the reserves the resources (allocation process) for the corresponding query (lines 8 to 16 in Alg.~\ref{alg:schedulingAgent}). For the allocation 
process, the Scheduling Agent verifies the system state and the SLA (see Alg.~\ref{alg:schedulingAgentState}). If the system state is $normal$ and the $SLA=acceptable$, it allocates the resources (lines 1 to 7 in Alg.~\ref{alg:schedulingAgentState}); if the system state is $degraded$, but the $SLA=acceptable$, it allocates the resources (lines 8 to 12 in  Alg.~\ref{alg:schedulingAgentState}); otherwise it applies its rules and acts according the available options ( lines 13 to 17 Alg.~\ref{alg:schedulingAgentState}).

\begin{algorithm}
\scriptsize
 \Input{(i), resources request; (ii), available resources}
 \Output{(i), allocated resources}
 \Switch{communication interface}{
  \Case{input interface}{
  	\If{resources request}{
        \textbf{check} available resources\;
        \textbf{send} request to the Controller Agent\;
    }
  }
   \Case{input interface}{
   		\If{analyzerProcess returns resources}{
        	TAB.resources = [null]\;
            \While{receiving response from the controller agent}{
            	TAB.resources = [CA1.resources, CA2.resources,...]\;
                \textbf{return} TAB.resources\;
                 \textbf{store} resources' data in the knowledge base\;
                 //allocation process\\
                 \If{available resources $\equiv$ needed resources}{
                 	
                    	Alg.~\ref{alg:schedulingAgentState}
                    
                 }
            }
        }
   }

 }
 \caption{Scheduling Agent (SA)}
 \label{alg:schedulingAgent}
  \vspace{-0.1cm}
\end{algorithm}

\begin{algorithm}
\scriptsize
\Switch{system state}{
\Case{normal}{
  \If{SLA $\equiv$ acceptable}{
  \textbf{allocate} resources\;
  \textbf{return} resources\;
  }
  }
  \Case{degraded}{
  \uIf{SLA $\equiv$ acceptable}{
  \textbf{allocate} resources\;
  \textbf{return} resources\;
  }
  \Else{
  available options = (retry or replacement or reallocation)\;
  \textbf{go to} line 7\;
  }
  }
  }
  \caption{Scheduling Agent (SA) state verification}
 \label{alg:schedulingAgentState}
\end{algorithm}

\vspace{-0.1cm}
\subsection{The Controller Agent}

The goal of the controller agent is to manage datacenter resources, by
tracking their status; through the Resource Status Updater component.
Figure~\ref{fig:CAArchitecture} illustrates its detailed architecture. The knowledge base of this agent contains the following information:

\begin{figure}
\centering
   \includegraphics[width=\columnwidth]{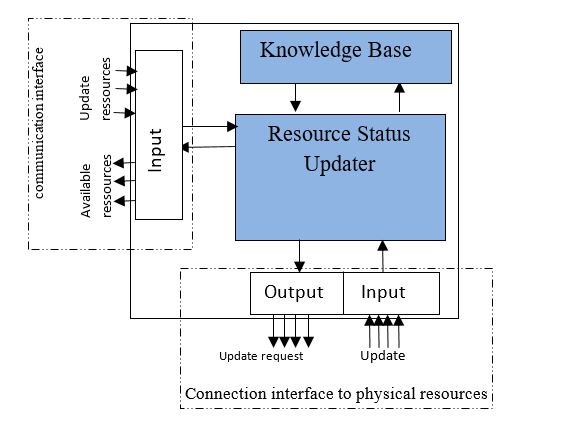}
   \vspace{-0.3cm}
   \caption{Controller Agent Architecture.}
   \label{fig:CAArchitecture}
   \vspace{-0.4cm}
\end{figure}

\noindent 
{\bf - Individual knowledge:} reflects the agent self-knowledge, including:
\begin{itemize}
\item \textit{name}: Controller Agent (CA);
\item \textit{address}: the location where it is deployed; e.g., physical servers in the data center;
\item \textit{individual objectives}: the objective of this agent is to track resource status;
\item \textit{state}: this agent can have the following execution states: receiving the request from the Scheduling  Agent and sending the available resources to Scheduling Agent.
\end{itemize}
\noindent
{\bf - Controller knowledge:}  it refers to the information this agent has about the status of the different resources in the datacenter.

Finally, its detailed behavior is illustrated in the algorithm presented in Alg.~\ref{alg:controllerAgent}. It permanently monitors the resources' states and updates this information in its knowledge base (lines 7 to 15 in Alg.~\ref{alg:controllerAgent}) . When a request from the Scheduling Agent arrives, it responds with the resources' statuses (line 18 in Alg.~\ref{alg:controllerAgent}). 

\begin{algorithm}
\scriptsize
 \Input{(i), Scheduling Agent's request}
 \Output{(i), available resources response}
 \Switch{communication interface}{
  \Case{input interface}{
  	\If{Scheduling Agent's request}{
        \textbf{update} resources\;
    }
  }
   \Case{physical output interface}{
   		\If{connection.resources}{
        	resources.stat = getNewStat(resources.stat)\;
            \textbf{return} resources.stat\;
        }
   }
    \Case{physical input interface}{
            \textbf{get} resources.stat\;
            \textbf{store} resources.stat's data in the knowledge base\;
   }
    \Case{output interface}{
            \textbf{send} resources.stat to Scheduling Agent\;
   }

 }
 \caption{Controller Agent (CA)}
 \label{alg:controllerAgent}
  \vspace{-0.1cm}
\end{algorithm}

\vspace{-0.1cm}
\subsection{Resource allocation and inter-agent Interaction}

\begin{figure*}
\centering
   \includegraphics[width=\textwidth]{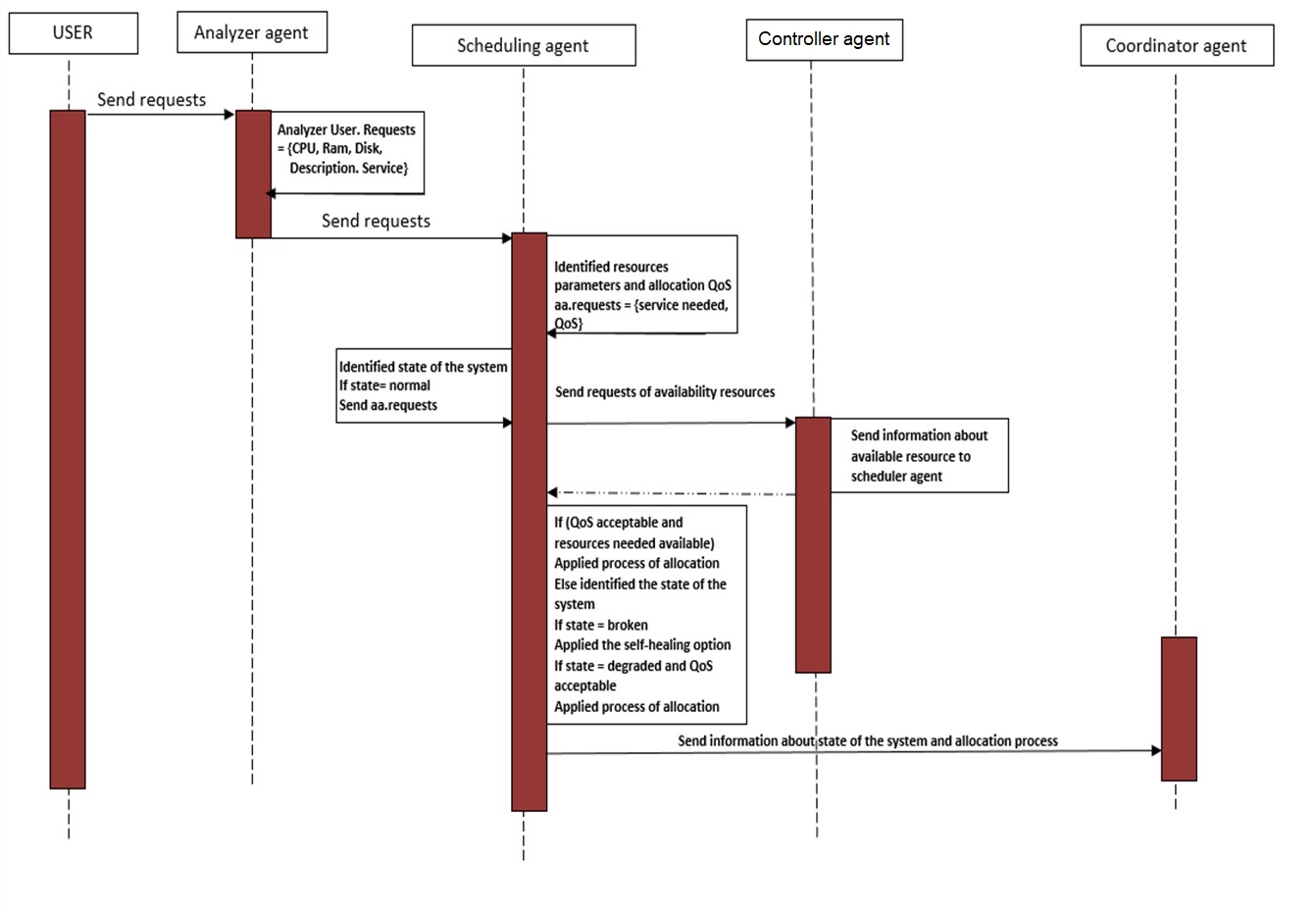}
   \caption{Inter-agent Interaction.}
   \label{fig:AgentInteractions}
\end{figure*}

The interaction among agents is illustrated in Figure~\ref{fig:AgentInteractions}. 
The interaction process starts when the Analyzer Agent receives a user request, which is analyzed, interpreted, sent it 
to the Scheduling Agent.
The Scheduling Agent verifies the system state (\texttt{state == normal}) and sends a request to the Controller Agent asking for information about the free resources. As a reminder, the Controller Agent tracks the status of resources in the datacenter.
When the Scheduling Agent gets the response from the Controller Agent, it verifies if the proposed QoS and available resources match the user requirements before making a final decision about resource allocation.
Finally, the Scheduling Agent sends a detailed report about the process and decisions made to the Coordinator Agent that supervises the whole process. The possible agent states are \textit{normal}, \textit{degraded}, and \textit{broken}. An agent is in the \textit{normal} state if the execution goes as expected concerning QoS allocation and system failures.
An agent is in the \textit{degraded} state if there is a degraded QoS allocation or failures in the system, but the QoS allocation still falls within the allowed range. Finally, an agent is in the \textit{broken} state if the QoS allocation is not acceptable or there are irreparable failures in the system.


\section{Illustrative Example}
\label{sec:architecture}





To illustrate our approach, we propose a scenario of resource allocation, shown in Figure~\ref{fig:allocationScenario}. The steps of the allocation process are the following:

\begin{figure*}
\centering
   \includegraphics[scale=0.72]{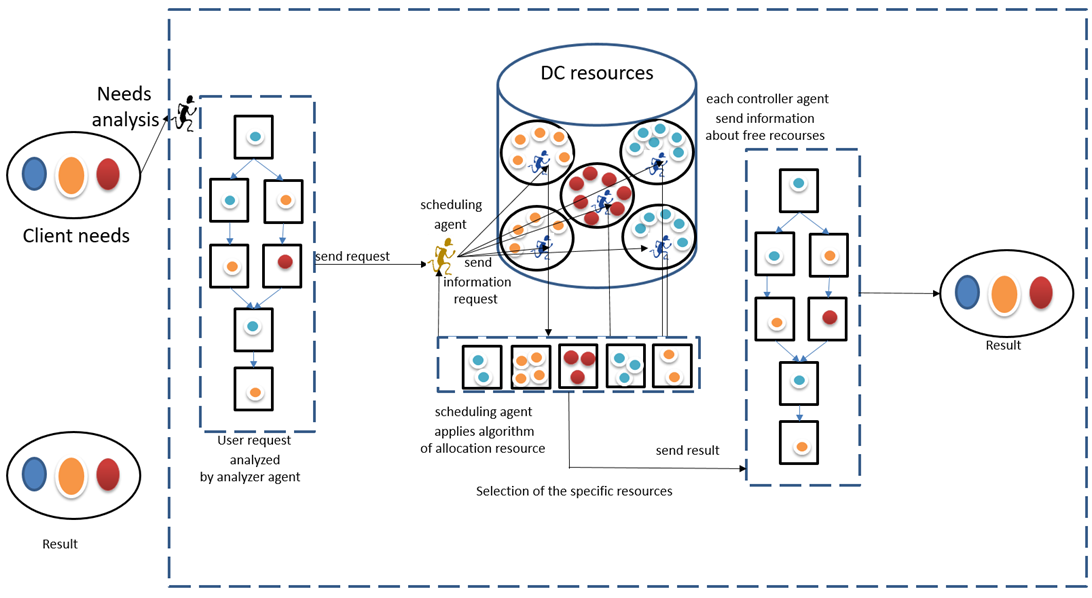}
   \caption{Allocation Scenario.}
   \label{fig:allocationScenario}
\end{figure*}

\begin{enumerate}
\item In the first step, a user requests the desired services and resources through a web interface, by expressing the services/resources needed and the expected QoS (e.g., desired resource capacity, acceptable range of capacity, acceptable time and price).  
The Analyzer Agent gets this request and studies the existing services/resources to extract the requirements concerning the needed resources and the expected QoS. Some examples of resources are CPU, RAM, and disk space. At the end of this step, the Analyzer Agent builds a query, called  $Q$, specifying the needed resources and expected QoS, and forwards it to the Scheduling Agent.
\item In the second step, the Scheduling Agent extracts the resource requirements forwarded by the Analyzer Agent. Its goal is to prepare the necessary resources to satisfy the request $Q$, by building a Virtual Machine, called it $VM_Q$.

The request $Q$ forwarded by the Analyzer Agent is composed of three elements, as follows $Q=(R,C,QoS)$, where:\\

\vspace{-0.2cm}
\noindent
$R=(r_1, r_2, r_3, ... , r_n)$ represents resources and services needed by the user;\\

\vspace{-0.2cm}
\noindent
$C=(c_1, c_2, c_3, ... , c_n)$ is the capacity corresponding to each resource; and\\

\vspace{-0.2cm}
\noindent
$QoS=(qos_1, qos_2, qos_3, ... , qos_n)$ represents the QoS for each resource/service.\\

\vspace{-0.1cm}
We represent the combination of resources and their corresponding requested capacities and QoS to build the suitable $VM_R$ as:

\vspace{-0.1cm}
\begin{equation*}
VM_R = \begin{pmatrix} 
r_1 & c_1 & qos_1\\ 
r_2 & c_2 & qos_2 \\ 
\vdots & \vdots\\
r_n & c_n  & qos_n
\end{pmatrix} 
\end{equation*}

\item After the identification of requested resources, the Scheduling Agent
sends a request to the Controller Agent to demand the corresponding free resources.

\item The Controller Agent receives the request from the Scheduling Agent and responds with the current state of the physical machines. There exists one Controller Agent, $CA_i$, for each physical machine, $PM_i$, in the Cloud. $CA_i$ manages the resources and information of its corresponding $PM_i$, including:

\begin{itemize}
\item The set of virtual machines that run in $PM_i$, denoted as
$V_i=(VM_1, VM_2, ... , VM_m)$;
\item The set of capacities corresponding to each virtual machine, denoted as $C_i=(c.VM_1, c.VM_2, ...,  c.VM_m)$;
\item The set of available resources on in $PM_i$, denoted as 
$RA_i=(ra_1, ra_2, ... , ra_p)$;
\item The set of capacities corresponding to each available resource, denotes as $CA_i=(ca_1, ca_2, ..., ca_p)$.
\end{itemize}

Then, the Controller Agent sends this information to the Scheduling Agent. We model the response of a Controller Agent $CA_i$ as two matrices:

{\scriptsize
\begin{align*}
&T.PM_i =
<
\begin{pmatrix} 
VM_1 & c.VM_1  \\ 
VM_2 & c.VM_2  \\ 
\vdots & \vdots\\
VM_n & c.VM_n  
\end{pmatrix} 
,
\begin{pmatrix} 
ra_1 & ca_1 \\ 
ra_2 & ca_2  \\ 
\vdots & \vdots\\
ra_p & ca_p  
\end{pmatrix} 
>
\end{align*}
}

\item In the next step, the Scheduling Agent stores the information received from the Controller Agent in its knowledge base; then, it builds a global matrix of available resources at time $t$, denoted as $A(t)$:

\begin{align*}
A(t) =
\begin{pmatrix} 
PM_1 & RA_1 & CA_1\\ 
PM_2 & RA_2 & CA_2 \\ 
\vdots \\
PM_n & RA_n & CA_n
\end{pmatrix} 
\end{align*}


where $PM_i$ are physical machines and $RA_i$ and $CA_i$ are the set of available resources with their corresponding capacities.

\item Finally, after building the global resource matrix, the Scheduling Agent applies its self-adaptive allocation algorithm to build from $A(t)$ a $VM$ that satisfies $VM_Q$.
\end{enumerate}







\section{Experiments}
\label{sec:exp}
We measured the performance of our multi-agent system in terms of  energy consumption, by simulation. To do so,
we used OMNeT~\cite{varga2008overview} and ICan Cloud~\cite{nunez2012icancloud} to simulate its functionalities. OMNeT is an extensible, modular, component-based C++ simulation library and framework, primarily for building network simulators. ICan Cloud is a simulation platform aimed to model and simulate Cloud computing systems. This simulation framework has been developed on the top of OMNeT++ and INET frameworks. 
The multi-agents system and the scheduling algorithm have been developed in C++, by using sockets to simulate their communication.

We simulated a scenario for allocating resources as mentioned in Section~\ref{sec:architecture}. Subsequently, we implemented a Cloud computing service provider to offers Cloud services to users. The data center implemented for this Cloud runs five heterogeneous servers. We randomly generated queries ($Q=(R,C,QoS)$) to submit to the Cloud in randomly time intervals (in seconds) and measured the aggregated energy consumption of the five servers. 



\begin{figure*}
\centering
   \includegraphics[scale=0.38]
   {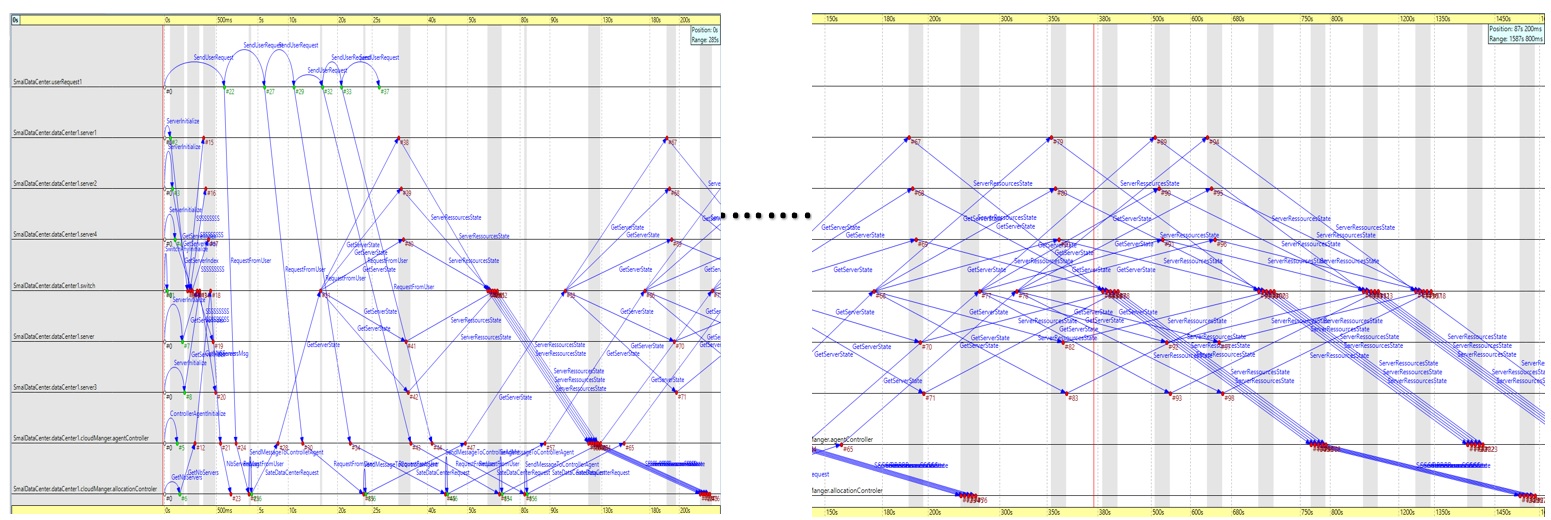}
   \caption{Extract of the Simulation Output.}
   \label{fig:simulation}
\end{figure*}

\begin{figure}
\centering
   \includegraphics[width=\columnwidth]{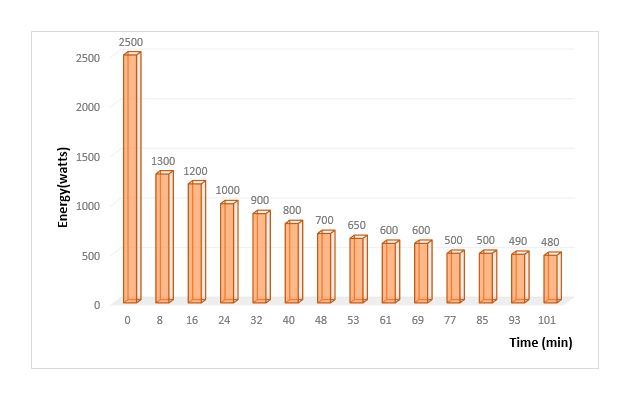}
   \caption{Energy consumption \textit{Server 1}.}
   \label{fig:results1}
\end{figure}

\begin{figure}
\centering
   \includegraphics[width=\columnwidth]{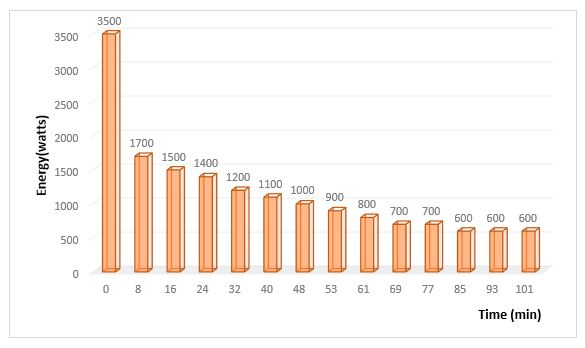}
   \caption{Energy consumption \textit{Server 2}.}
   \label{fig:results2}
\end{figure}

Figure~\ref{fig:simulation} shows an extract of the simulation running. Note that queries are submitted at different time intervals: the first one at time 0, the second one at 500ms, the third one at 5sec, and so on until the end of the simulation at 102min.
All queries were satisfied in terms of needed resources and required QoS. 
Figures~\ref{fig:results1} and~\ref{fig:results2} show the energy consumption rates in Server 1 and Server 2, respectively. 
When we apply our self-adaptive strategy for resource allocation, we observe a remarkable change in the level of energy consumption, showing the efficiency of the scheduling algorithm. 
For Server 1 (Figure~\ref{fig:results1}), the decrease of energy consumption was from 2500 megawatts to 480 megawatts, while for Server 2 (Figure~\ref{fig:results2}), it was from 3500 megawatts to 600 megawatts. These decreases of energy consumption happen at times ranging from 0 minutes to 102 minutes.

These results put in evidence the performance of our algorithm of QoS allocation, confirming the choice of using a multi-agent system implementing a self-adaptive mechanism for the smart management of resources in Cloud data centers.

\section{Conclusions}
\label{sec:conclusions}

In this work, we tackled the problem of resource allocation in Cloud platforms. To that end, we have proposed a multi-agent architecture based on the {\it MAPE-K} model. For our future work, we plan to present a formal model of our approach defining the query for resource allocation, user preferences, agent rules, and state transitions. We also plan to describe in detail the scheduling algorithm and present additional experiments  to evaluate other performance aspects, besides energy consumption.

\bibliographystyle{plain}
\bibliography{sample}

\end{document}